\documentclass[aps,prb,groupedaddress]{revtex4}
\usepackage{graphicx}
\begin{document}
\title{Collapse of  the charge ordering state at high magnetic fields in the rare-earth manganite, Pr$_{0.63}$Ca$_{0.37}$MnO$_3$}

\author{K. S. Nagapriya\footnote[1]{Electronic mail: ksnaga@physics.iisc.ernet.in}, A. K. Raychaudhuri\footnote[2]{Electronic mail: arup@physics.iisc.ernet.in}, Bhavtosh Bansal, V. Venkataraman}
\address{Department of Physics, Indian Institute of Science,  Bangalore 560 012,  India.}
\author{Sachin Parashar,  C. N. R. Rao}
\address{Jawaharlal Nehru Center for Advanced Scientific Research, Bangalore , India}
\date{\today}
\begin{abstract}
We have investigated the specific heat and resistivity of a single crystal of Pr$_{0.63}$Ca$_{0.37}$MnO$_3$ around the charge ordering (CO) transition temperature, $T_{CO}$, in the presence  of high magnetic fields ($\leq 12T$) which can melt the charge ordered state.   At low magnetic fields ($\leq 10T$),  the manganite transforms from a charge-disordered paramagnetic insulating (PI) state to a charge-ordered insulating (COI) state as the temperature is lowered. The COI state becomes unstable beyond a threshold magnetic field and melts to a ferromagnetic metallic phase (FMM). This occurs for $T < T_{CO}$. However, above a  critical field $\mu_0H_{\rho}^*$, the sample shows the onset of a metallic phase for $T > T_{CO}$ and the COI transition occurs from a metallic phase. The onset temperature of the high-field metallic behavior  decreases with an increase in the field and above a field $\mu_0H^*$, the COI transition does not occur and the CO state ceases to occur at all T. The entropy change involved in the CO transition, $\Delta S_{CO}$ $\approx$ 1.6J/molK at 0T, decreases with increasing field and eventually vanishes for a field $\mu_0H^*$. The collapse of the CO state above $\mu_0$H$^*$ is thus associated with a collapse of the entropy that stabilizes the CO state.  

\noindent 
\end{abstract}
\maketitle
\section{Introduction}
Thermodynamic and transport properties of rare-earth manganites with the general formula R$_{1-x}$A$_x$MnO$_3$ (R:  a trivalent rare-earth ion and A: a divalent alkaline-earth ion) have attracted considerable current interest.~\cite{i2,i3} For certain values of x, the Mn$^{3+}$ and Mn$^{4+}$ order in the lattice giving rise to what is called charge ordering. The CO insulating (COI) state can be destabilized by various perturbations such as temperature, magnetic field, electric field and radiation  to a ferromagnetic metallic (FMM) or a charge disordered and paramagnetic insulating (PI) state~\cite{d1,d2,d3,d4,Ayanrt,prna}. 
\begin{figure}
\includegraphics[width=8cm]{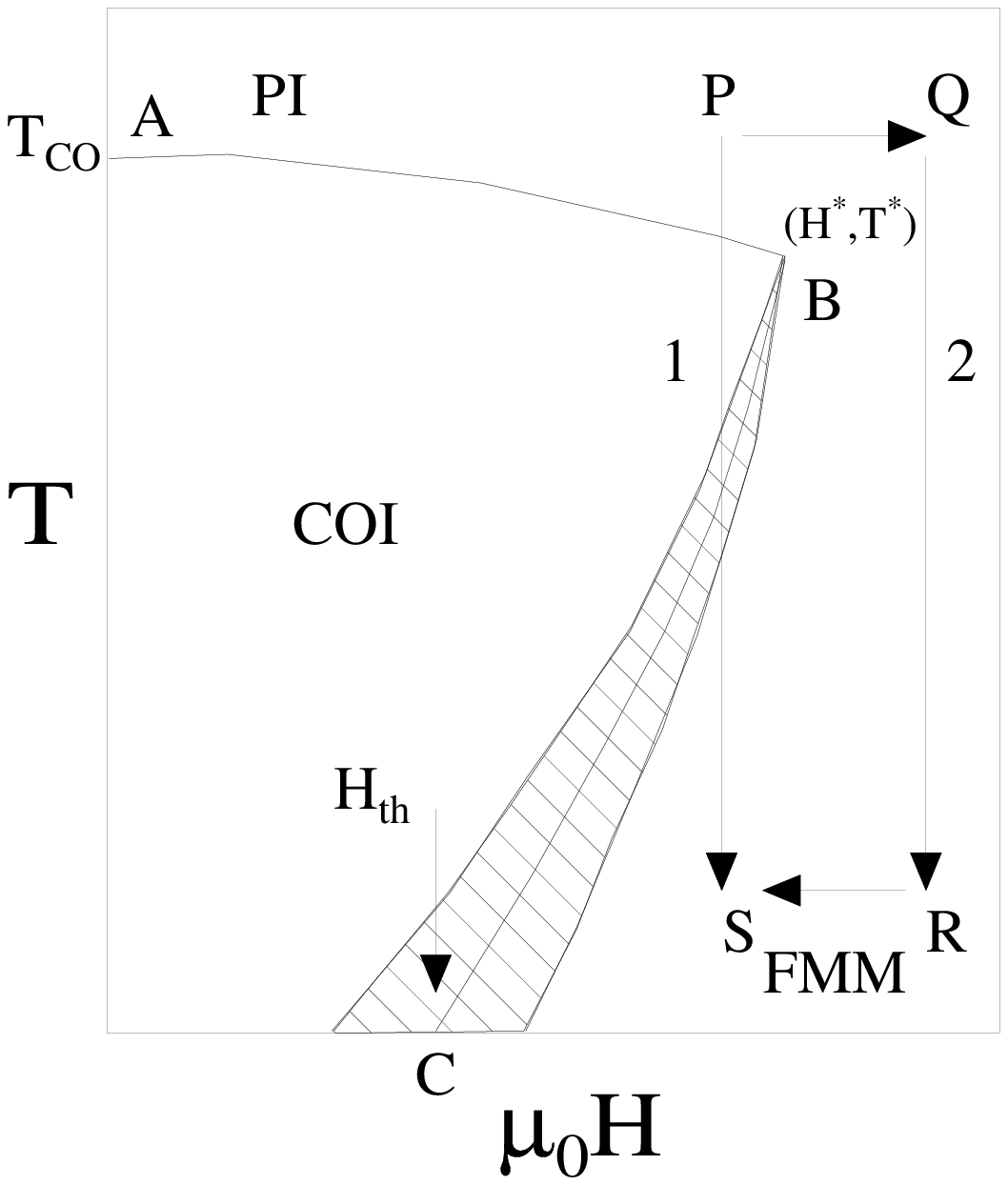}
\caption{\label{fig:figure1} A cartoon  of the T-$\mu_0$H plane phase diagram seen in most manganites like Pr$_{1-x}$Ca$_x$MnO$_3$ which show the onset of CO transition from a PI (charge disordered) state. COI is charge ordered insulator and FMM is ferromagnetic metal.}
\end{figure}

Of interest to us is the observation that an applied magnetic field can destabilize the charge-ordered state leading to a ferromagnetic metallic state. This  phenomenon has been studied in detail by using various techniques.~\cite{Ayanrt,Teresa,Hardy,Smol} It appears that there is a minimum threshold field $\mu_0H_{th}$ which is needed to melt the CO state to a FMM phase below $T_{CO}$. $\mu_0H_{th}$ depends on a number of factors like the radius of the A site cation, $<r_A>$ (equivalently the bandwidth), and deviation from the x=0.5 composition which has equal amount of carriers and holes. $\mu_0H_{th}$ decreases as $<r_A>$ increases and we move away from the x=0.5 filling~\cite{i3}. 
In our present investigation we explore whether there is an upper field  $\mu _0H^*$ ($\mu _0H^* > \mu _0H_{th}$) beyond which the CO transition cannot occur at any temperature (i.e., the temperature $T_{CO}$ ceases to exist). We then investigate the region of phase diagram close to $\mu _0H^*$ and beyond. We note that on comparison of various data available in different CO systems it appears  that there is a likelihood of such an upper field as $\mu _0H^*$~\cite{Ayanrt,Teresa,Hardy,Smol}. The present investigation goes beyond these likely evidences and establishes concretely existence of such a field.
\begin{figure}
\includegraphics[width=8cm]{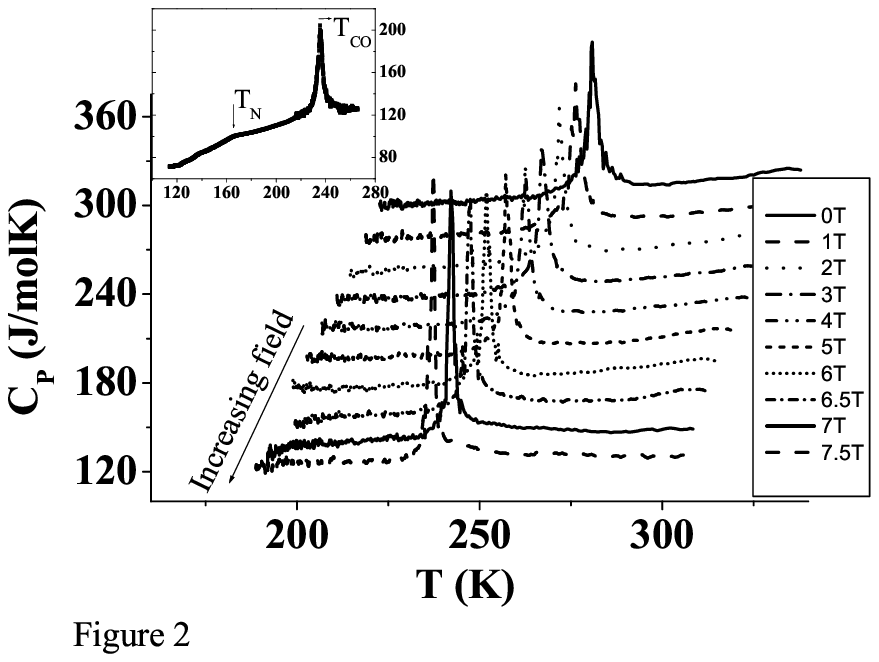}
\caption{\label{fig:figure2}Waterfall plot of specific heat as a function of temperature at various magnetic fields. Note that the data are offset for clarity Inset: Specific heat as a function of temperature for zero  field over an extended temperature range. Note the T$_{CO}$ at $\approx$ 235K and T$_N$ at $\approx$ 165K.}
\end{figure}

The cartoon in figure~1 explains the issues that we are probing in this investigation. The cartoon is representative of the currently accepted ($T-\mu_0H$) plane phase diagram seen in  manganites such as Pr$_{1-x}$Ca$_x$MnO$_3$,  showing the onset of CO transition from a paramagnetic insulating (PI) that is charge disordered state. The CO transition occurs at zero field at the temperature $T_{CO}$, marked by point A. On the application of a magnetic field, $T_{CO}$ generally decreases and follows the line AB. Thus PI $\rightarrow$ COI transition occurs when the line AB is crossed. However, for $\mu_0H >$ $\mu_0H_{th}$, the COI state melts to FMM state at $T < T_{CO}$, as mentioned before. The melting of the COI state to the FMM state occurs along the line BC which also shows hysteresis and often strong time dependent behavior~\cite{Anane}. The low temperature COI $\rightarrow$ FMM transition occurs on crossing line BC. The region bounded by ABC is a region of "mixed phase" where two phases can co-exist. The transition across the line AB has a clear thermodynamic signature and is associated with a change in entropy~\cite{Ayansph}. In the case of Pr$_{0.63}$Ca$_{0.37}$MnO$_3$ ($T_{CO}$ $\approx$ 235K), the entropy removed at $T_{CO}$ at zero magnetic field is $\approx$ 1.6J/molK as found from direct measurements of specific heat~\cite{Ayansph}. A good part of the entropy change is released as latent heat and the transition is believed to be of first order. There is no large thermodynamic signature to the transition across the line CB when the COI melts in a magnetic field, although the existence of hysteresis (marked by hatches) as well as the suddenness of the transition is often interpreted as a first order transition. Whatever be the case, there is very little change in the specific heat as well as entropy across the line BC. The field $\mu_0$H$^*$ refers to the point B where the two lines meet at temperature $T^*$. The above phase-diagram though known and well established is presented here to put this investigation in proper perspective.

The present investigation is primarily focused on the region around the point B ($\mu_0H^*, T^*$). The primary motivation is to explore the following issues:
\begin{enumerate}
\item{The ($\mu_0H-T$) phase diagram as depicted in figure~1 is expected to have more features. This can be illustrated as follows. If we cool down following path 1, we start from a PI phase at point P, cross the boundaries AB and BC and terminate in a FMM phase at point S. But if we follow the path 2 (PQRS), we can go from P (PI phase) to S (FMM phase) without any phase transition or cross-over region. This implies that somewhere along path 2 there should be a boundary demarcating two phases or a cross-over. This should occur for $\mu_0H >$ $\mu_0H^*$. We note that this issue has not been raised in previous publications in this field in the context of Pr$_{0.63}$Ca$_{0.37}$MnO$_3$, although the  phase diagram in this field region where the CO transition ceases to exist has been discussed  in a related though different system Pr$_{0.75}$Na$_{0.25}$MnO$_3$ which we will discuss in appropriate place. }
\item{Our previous studies of specific heat at $\mu_0H$ = 0 and $\mu_0H$ = 8T have shown that in the transition across the line AB, there is a finite entropy change ($\Delta S_{CO}$ $\neq$ 0)~\cite{Ayansph}. However, there is a negligible change is entropy as one crosses the line CB. One would therefore expect that the nature of the transition changes as one goes along the line AB. We explore the evolution of $\Delta S_{CO}$ as we approach the point B. Does $\Delta S_{CO} \rightarrow$ 0 as the point B is approached?}
\item{At fields lower than $\mu_0H^*$, the  high temperature PI phase makes a transition to the COI phase at $T_{CO}$  which eventually melts at an even lower temperature in a magnetic field, giving rise to the FMM phase. Is there a change associated with the PI phase in a high enough magnetic field? This particular aspect has also not been explicitly discussed/observed before although certain past studies show likely signs of such metallic phases~\cite{prb1996}.}
\end{enumerate}
\begin{figure}
\includegraphics[width=8cm]{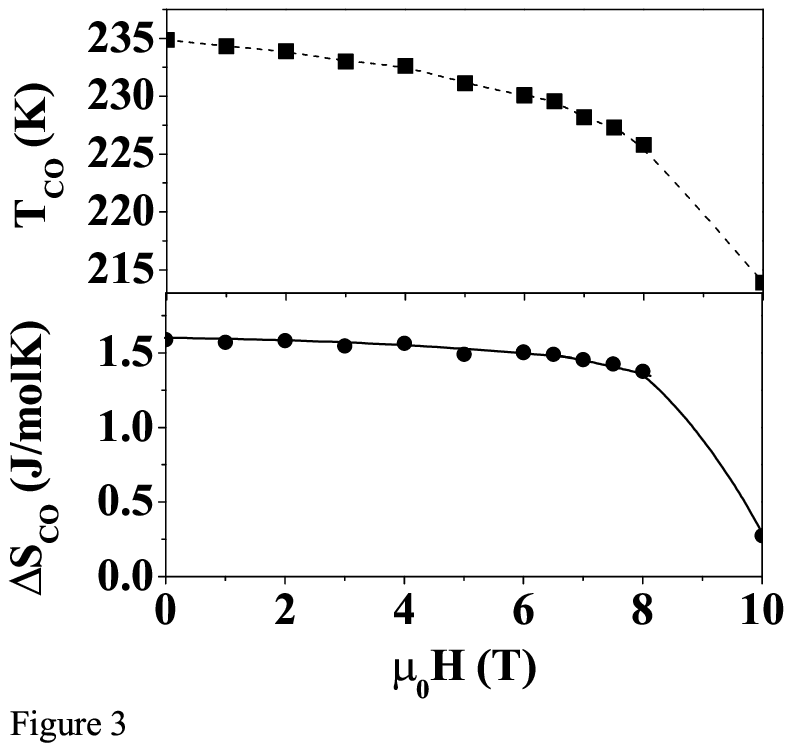}
\caption{\label{fig:figure3}(a) T$_{CO}$ as a function of applied magnetic field. (b) Change in entropy at the CO transition as a function of field.}
\end{figure}

We investigate the above issues in a single crystal of Pr$_{0.63}$Ca$_{0.37}$MnO$_3$ by using both thermodynamic (specific heat) and transport measurements in a magnetic field upto 12T. The results of the investigation (discussed in detail in subsequent sections) establish
\begin{enumerate}
\item{As we move along the line AB, as $T \rightarrow T^*$ and $\mu_0H \rightarrow$ $\mu_0H^*$, $\Delta S_{CO}$ $\rightarrow$ 0. Beyond this field,  CO does not exist at any temperature. The $T_{CO}$ itself decreases as H increases. The fact that no CO transition takes place when $\Delta S \rightarrow 0$ means that there is an essential role of entropy in stabilizing the CO state.}
\item{In the temperature range ($260K >T > T_{CO}(H)$), the PI phase crosses over to a metallic phase for a certain magnetic field. We designate this regime as a  metallic regime (M). In this field and temperature regime, the CO transition occurs from a metallic phase, instead of a insulating phase. This gives us a new line of cross-over in the ($T-\mu_0H$) phase diagram. Above 260K the resistivity reaches a rather field insensitive region.}
\item{For fields $\mu_0H > \mu_0H^*$ (which is the limit of stability of CO transition) there appears to be the presence of a cross-over region (or phase transition) within the metallic phase as it is cooled where long range ferromagnetic order sets in (FMM) at $T \approx$ 220K.}
\end{enumerate}

\begin{figure}
\includegraphics[width=8cm]{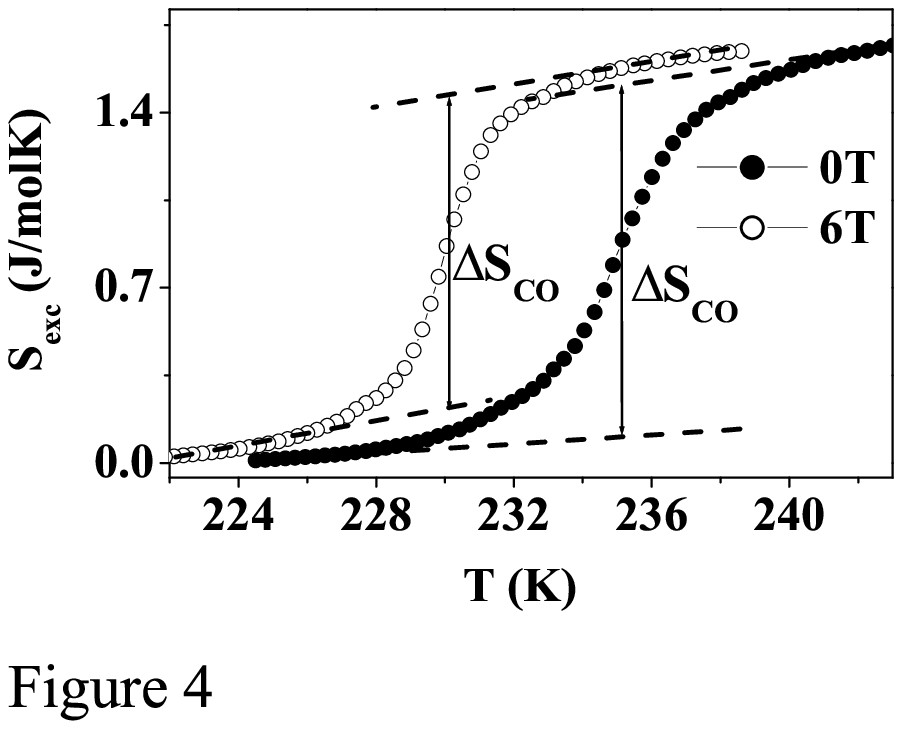}
\caption{\label{fig:figure4}S$_{exc}$ as a function of temperature for fields of 0T and 6T.}
\end{figure}

To our knowledge, the studies reported by us are  new and they make the qualitative phase diagram  in figure~1 richer and more complete, particularly in the high field region.

\section{Experimental techniques}
\subsection{Sample}
The sample we have chosen for the present investigation Pr$_{0.63}$Ca$_{0.37}$MnO$_3$ is a well characterized single crystal,  with a  well-defined CO transition at $T_{CO}$ $\approx$ 235K. The crystal has been grown by float-zone technique using a mirror furnace. This material has been extensively investigated by us. 
Previously reported~\cite{Ayansph} specific heat measurements on this sample from our group showed that the CO transition was most likely first order in nature and the entropy change involved in the transition had been calculated. We note that in the past investigation from our group~\cite{Ayansph} the measurements were not done to high enough magnetic field so that one can reach the region ($T^*,H^*$) and beyond. As a result observations made in the present paper could not be made. In particular we could not observe $\Delta S \rightarrow 0$. Also no concrete connection to transport studies at high fields had been established.
\subsection{Measurement of Specific heat in magnetic field}
The specific heat was measured using the technique of continuous cooling calorimetry~\cite{Raj}. This is distinct from our studies in reference ~\cite{Ayansph} where data were taken by adiabatic calorimetry. The temperature range in which the calorimeter  operates is from 100K-400K (this is the temperature range in which the transitions occur).  The specific heat in this temperature region was measured in a field upto 10T. The calorimeter can measure heat capacity of very small samples ( mass $\sim$ mg. The addenda heat capacity is $<$ 20 mJ/K at room temperature. 

\begin{figure}
\includegraphics[width=8cm]{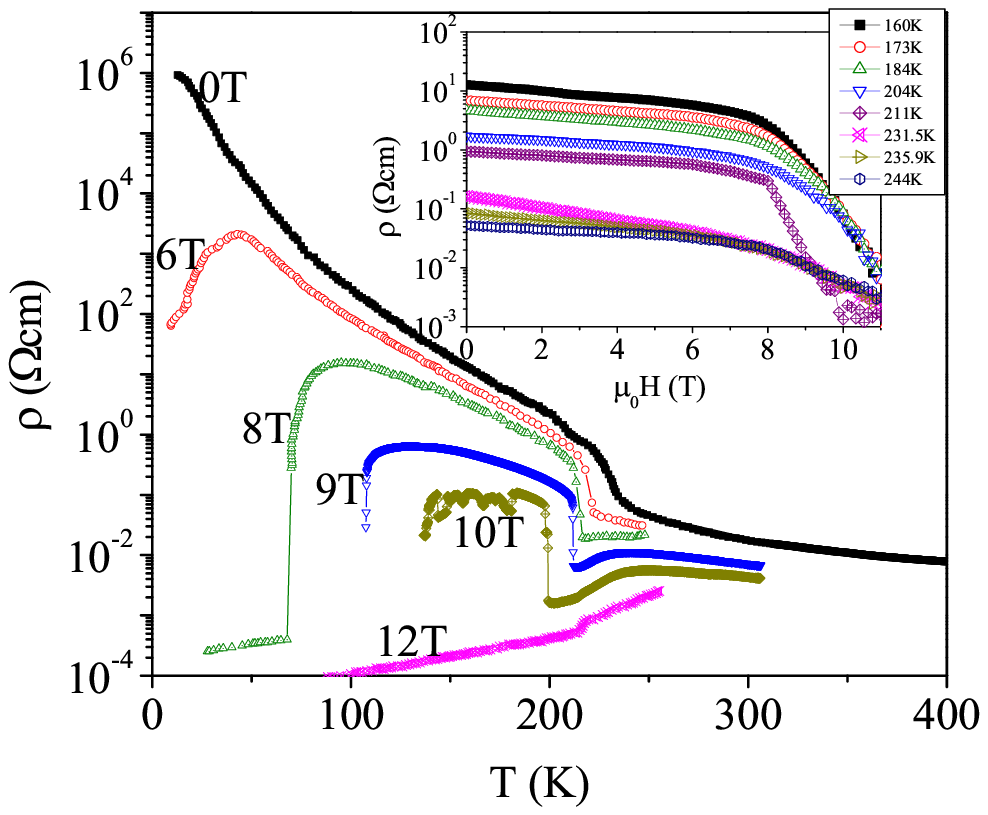}
\caption{\label{fig:figure5}Resistivity as a function of temperature for a few representative fields. Inset shows resistivity as a function of field at different temperatures.}
\end{figure}

In this method, the sample temperature is recorded as a function of time as the sample (warmed upto a predetermined temperature by a heater) cools continuously, losing heat to the base through a heat link  whose characteristics are experimentally determined. The cooling curve is determined by the equation
\begin{equation}
C(T)\frac{dT}{dt}(T) = -P_{leak}(T)
\end{equation}
where $P_{leak}(T)$ is the power leaked to the bath ($P_{leak}(T)$ is experimentally determined and has both the conduction and radiation contributions). A platinum film on the substrate works as both a heater and thermometer. The specific heat and latent heat at a transition are deduced from the cooling curve - by determining $dT/dt(T)$ and the thermal link characteristics. The data obtained by this method matches with that obtained by reference ~\cite{Ayansph} which was taken by adiabatic calorimetry which acquires data during heating. There are some differences in the transition region that we discuss. This difference shows up in the transition region as a small change in the width of the transition as well as height of the heat capacity peak. This of course depends on the rate at which temperature changes in this region. However, this does not have much effect when we integrate the excess heat capacity at transition to find entropy. The uncertainty in entropy due to the experimental contributions are limited to within 10$\%$. 

\subsection{Measurement of resistivity in steady and pulsed magnetic fields}                                
We measured the resistivity  using a standard four probe method. For making the contacts, four gold contact pads were evaporated on the sample and the contacts to the sample were subsequently made by soldering 40 $\mu$m copper wires using Ag-In solder.  Most of the MR data were taken by a superconducting solenoid producing steady field upto 12T. For comparison, we have also used a pulsed magnet to make the resistivity measurements. The pulsed field with $\mu_0$H$_{peak}$ = 14T has a fall time 20msec. The data acquisition was done with a 12 bit 20MHz card.  In case of the superconducting magnet the data were taken by fixing the field and varying the temperature. In case of the pulsed magnet it was field scans at fixed temperatures. Since we are in a regime where there is no noticeable hysteresis effects both the methods should lead to the same result provided the field and temperature calibration has the same traceability. We find that the data taken by  the steady field and the pulsed field are close within experimental uncertainty. In the results presented we do not distinguish the data taken by the two methods and present them together.
\section{Results}
The results are presented in three sub-sections. In the first sub-section, we discuss the specific heat as a function of temperature at different magnetic fields. The second sub-section contains results of the entropy changes at the CO transition at different fields. Finally, we present the results obtained from  resistance measurements at different fields and temperatures.
\subsection{Specific heat as a function of temperature and magnetic field}
A waterfall plot of specific heat as a function of temperature at different magnetic fields is shown in figure~2 (The data are offset for clarity). In zero field, the sample shows a first order CO transition at $T_{CO}$ = 235K and a small step at $\approx$ 165K which is the Neel temperature $T_N$.  The zero field specific heat data are shown as inset in figure~2 for an extended temperature scale. In the specific heat data with increasing magnetic field, 
the $T_{CO}$ shifts to lower temperatures.
A plot of $T_{CO}$ as a function of the field is shown in figure~3. We see that a  field of 10T shifts the $T_{CO}$ by as much as 20K.   
 It can be seen from figure~3 that there are two regions. For $\mu_0H <$ 6T, the change in  $T_{CO}$ is gradual while for $\mu_0H>$ 6T, it is rather rapid. Close to 10T we see the transition as just  a sharp line and essentially no width as limited by the experimental set up. We show below that with increasing field, we ultimately reach a point where the entropy change at $T_{CO}$ vanishes. 
\subsection{Entropy change at $T_{CO}$}
It has been established by calorimetric investigations that there is a change in entropy $\Delta S_{CO}$ at the charge ordering transition. In zero field, $\Delta S_{CO}$ $\approx$ 1.6J/molK~\cite{Ayansph}. The earlier data were obtained by conventional adiabatic calorimetry wherein the data are taken while heating while the present data were taken by continuous cooling calorimetry wherein the data are taken during cooling.  Comparing the peak height or width taken by these two completely different methods may be difficult as there is an inherent error present in determining the peak width in first order transitions. What can be compared is the entropy change at the transition $\Delta S_{CO}$ as this is the area under the peak in specific heat divided by the temperature. This $\Delta S_{CO}$ for both data match to within experimental uncertainties.
For calculating the entropy change during the CO transition, a background lattice contribution has to be first subtracted out from the specific heat data. We define $C_{exc} = C - C_{lattice}$. For estimation of $C_{lattice}$ we employ the procedure described in detail in reference~\cite{Ayansph}.   The entropy change is calculated by numerically integrating $C_{exc}/T$ as $S_{exc}$ = $\int_{200}^T(C_{exc}/T)dT$. The lower limit of integration (200K) is so chosen that $C_{exc}$ $\cong$ 0 for all $\mu_0H$ at this temperature. Figure~4 shows the $S_{exc}$ as a function of temperature near $T_{CO}$. To obtain the change in entropy $\Delta S_{CO}$ at the transition we linearly extrapolate the values of $S_{exc}$ above and below the transition and find the difference in the extrapolated values at the transition. By this procedure we obtain a $\Delta S_{CO}$ of $\approx$ 1.6J/molK at 0T. (Note that this procedure does introduce uncertainty in the absolute value of $\Delta S_{CO}$ which we estimate $\simeq$ 10$\%$.)

\begin{figure}
\includegraphics[width=8cm]{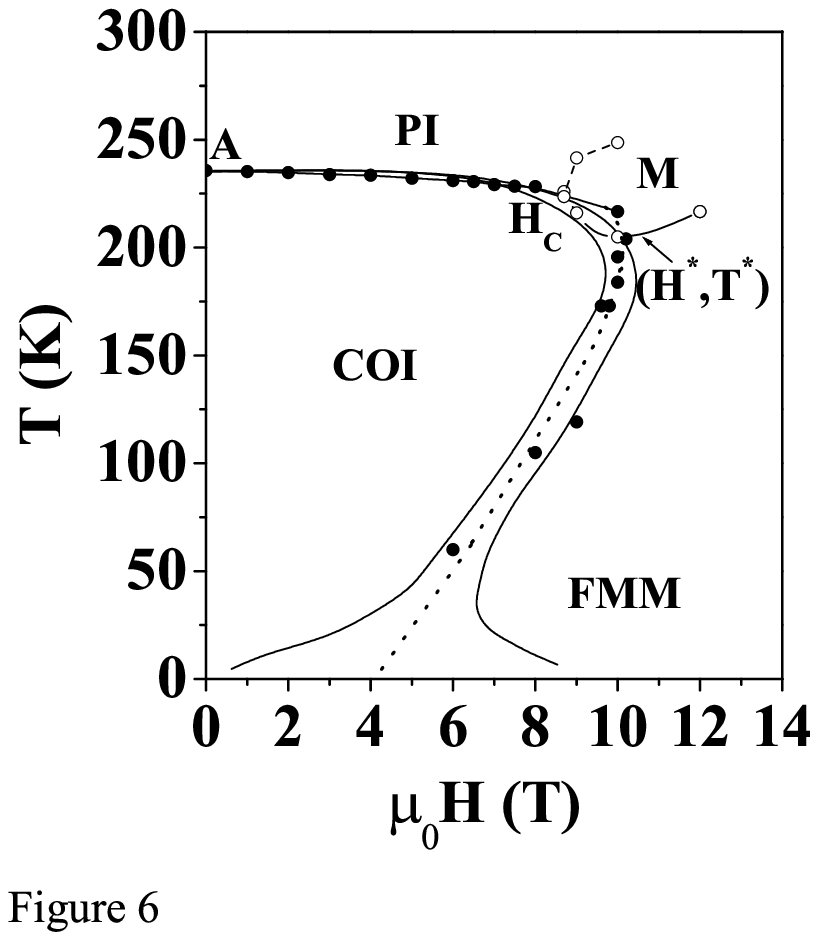}
\caption{\label{fig:figure6}T-$\mu_0$H plane phase diagram of Pr$_{0.63}$Ca$_{0.37}$MnO$_3$ combining the results of the present investigation and the existing phase diagram (solid line). M is metallic phase.}
\end{figure}

A plot of the calculated $\Delta S_{CO}$  as a function of field is shown in figure~3. 
The decrease in $\Delta S_{CO}$ is small upto a field of 6T. Above this, the fall is very rapid.  The important fact that we have established is that $\Delta S_{CO} \rightarrow 0$ as the field is increased. From our calorimetry data, we could clearly identify the point B as $T^*$ $\approx$ 215K, $\mu_0H^*$ $\approx$ 10.5T where $\Delta S_{CO} \approx 0$. Our transport measurements showed that at $\mu_0H^*$ the CO transition ceases to exist.

As pointed out earlier, no marked signature except a small feature is observable at low T as we cross the line CB. This becomes weaker as $\mu_0H$ increases. Thus vanishing of $\Delta S_{CO}$ as we move along the line AB is compatible with the fact that the two lines of transition meet at the point B where $\Delta S_{CO}$ $\approx$ 0. We point out that vanishing of the entropy $\Delta S_{CO}$ at a certain field is a new observation and we explain below that it is linked to the stability of the CO state.
\subsection{Resistivity and metal - insulator transition close to T$_{CO}$ at high fields}
In this sub-section, we present the results of the resistivity ($\rho$) measurements. The important observation that we make is that close to $T_{CO}$ (T $>$ $T_{CO}$),  the magnetic field induces a metallic behavior ($d\rho/dT > 0$) for $\mu_0H$ greater than  a certain  field $\mu_0H_{\rho}^*$ $\approx$ 8.7T.  For $\mu_0 H < \mu_0 H_{\rho}^* < \mu_0 H^*$, the CO transition occurs although it is from a  metallic phase. This occurs over a narrow range of field and temperature. For  $\mu_0 H > \mu_0 H^*$,  we find that there is a transition/cross-over to a metallic state and there is no signature of formation of an insulating state in resistivity. This establishes a close link between the calorimetry and transport data.

In figure~5 we show the $\rho$ vs $T$ plot at a few representative fields.  The inset shows $\rho$ vs $\mu _0H$ at different temperatures close to  $T_{CO}$. One can clearly see a strong negative magnetoresistance at high fields.  For T $<$ 235K,  the change in $\rho$ in a magnetic field is rather sharp.  In this region $T < T_{CO}$, and the drop in $\rho$ is melting of CO state to FMM state. It can also be seen that for fields  $\mu_0H$ $\geq$ 9T there is a temperature range where the resistivity  is  much like a metallic sample with $d\rho /dT >$ 0. This cross-over to a metallic behavior at high fields can be seen in figure~5. In the temperature regime $T > T_{CO}$ at low magnetic fields, $\rho$ has a negative temperature coefficient of resistivity $d\rho/dT < 0$ which is characteristic of an insulator. As the magnetic field is increased,  there is a gradual change in slope of $\rho$ vs $T$ curve. Beyond a field of $\mu_0H_{\rho}^*$ = 8.7T, $\rho$  develops a positive temperature coefficient of resistivity for $T$ close to $T_{CO}$. As it is cooled below $T_{CO}$,  the CO transition takes place and $d\rho /dT$ changes sign.  The occurance of the metallic state for field $\>\mu_0H_{\rho}^*$ may have its stability only over a limited temperature range. This is because for $T > 260K$ $d\rho /dT$ gradually decreases and the resistivity becomes insensitive to field. It may be that at higher temperature $\rho$  joins a commom curve with rather small  $d\rho /dT$. (Due to technical difficulties with the cryogen free magnet we cannot reach a temperature much higher than $270K-275K$ with a field of 12T This prevented us to make a definitive statement at high field and high temperature region.) However, it is clear from the data (figure~5) that in this field regime the COI transition takes place on cooling from a metallic state, unlike the CO transition at lower field where it takes place from an insulating phase. In this context it is interesting to recall that in the manganite system like $Nd_{0.5}Sr_{0.5}MnO_3$, which has somewhat larger bandwidth~\cite{i3}, the transition to COI state occurs from a metallic ferromagnetic state. In manganites with lower bandwidth the transition to COI state is always from an insulating state that is charge disordered.

To summarize, we find that the application of high magnetic fields causes significant changes in the material, particularly at $T \approx T_{CO}$. This is in addition to the field induced suppression of CO at lower temperatures. A combination of the transport and the calorimetric measurements   show that the COI state which is the ground state in zero field is thermodynamically unstable for $\mu_0H >$ $\mu_0H^*$ and ceases to exist. For $\mu_0H >$ $\mu_0H^*$ the COI state does not from at any T. There is a range of field 8.7T $\leq \mu_0H \leq$ 10.5T where $d\rho/dT >$ 0 for $T > T_{CO}$ so that  the COI transition occurs from a metallic (M) phase. At higher field and low temperature this metallic phase becomes ferromagnetic although at higher ($T>260K$) the field induced metallic phase appears to give way to an insulating state.

\section{Discussion}
In figure~6 we present the results of our experimental investigations in the form of a phase diagram. In the same figure,  we show the  phase diagram  constructed from data of previous experiments by  solid lines. The present experiments identify the new phases or cross-over curves. There are a number of issues in the phase diagram that need attention that have not been discussed before~\cite{note1}.

As noted  in our earlier publications,~\cite{Ayansph}  as we cross the line BC (see figure~1) there is a small thermodynamic signature of the transition with a heat release of  $\approx$ 10J/mol. This is much smaller than the $\Delta S_{CO}$ observed at the CO transition as we cross the line AB. The vanishing of $\Delta S_{CO}$ at the point B is thus internally consistent. The COI and FMM phases separated by the line BC are expected to be close energetically (similar heat capacity), and they also have similar entropy. The high-temperature charge disordered insulating phase (PI) has a higher entropy that gets released as $\Delta S_{CO}$ as we cross AB. Vanishing of $\Delta S_{CO}$ as the point B is approached implies that the three phases (PI, FMM and COI) have the same entropy at point B. 
It appears that the first order transition across AB ceases at B. This also makes the point B a special point in the phase diagram.

For $\mu_0H > \mu_0H^*$, the COI state  ceases to exist.  The COI phase, due to charge (and possibly orbital) ordering has a lower entropy in fields upto $\mu_0H <$ $\mu_0H^*$. The release of entropy at $T_{CO}$ stabilizes this phase. The magnetic field destabilizes this state making $\Delta S_{CO}$ $\rightarrow$ 0 as $\mu_0H$ $\rightarrow$ $\mu_0H^*$. In manganites with relatively broader bandwidths as in La$_{1-x}$Sr$_x$MnO$_3$ (x $\simeq$ 0.3-0.5), one finds such a PMM phase above $T_C$ at zero field~\cite{i3}.  For most manganites, with smaller bandwidths as in La$_{1-x}$Ca$_x$MnO$_3$ and Pr$_{1-x}$Ca$_x$MnO$_3$, the high-temperature phase is an insulating phase.  The observation of a  field induced cross-over to a metallic phase ($d\rho /dT >$ 0) although in a narrow temperature range ($260K >T > T_{CO}$) is therefore an interesting phenomenon. 

The insulating phase above $T_{CO}$, like other colossal magnetoresistive (CMR) systems is  attributed to Jahn-Teller (JT) distortion. In the undoped parent compound such as PrMnO$_3$ or LaMnO$_3$, the high-temperature JT distortion can be co-operative in nature leading to orbital ordering. This leads to an insulating state in these samples. Hole-doping (by substitution of a bivalent cation in the rare-earth site) leads to a dilution of the co-operative JT and orbital ordering although the high-temperature insulating phase is retained. Past studies (using resonant X-ray scattering technique or neutron scattering technique~\cite{neutscat}) have shown some degree of orbital or charge order above $T_{CO}$, albeit with a  small correlation length. The suppression of the insulating phase at $T > T_{CO}$ , would mean that a high enough magnetic field can suppress these local ordering leading to a metallic state with $d\rho /dT >$ 0 ,although in a limited temperature range. An interesting possibility is that the magnetic field increases the width of the conduction band and thus suppress polaronic effects and local orbital ordering. It is possible to model the behavior as co-existance of two phases (both paramagnetic) one metallic and the other insulating. The application of a magnetic field  increases the volume fraction of the metallic phase. The increase in volume fraction beyond a certain value leads to the cross-over to the metallic state. However,at high enough temperature the degree of spin ordering even at high  field may not be large enough to cause any effect on the polaronic or local orbital ordering thus stabilizing the insulating state.The metallic and inulating phases are identified as a change in $d\rho/dT$, occurs when the magnetic field reduces the resistivity to about 20m$\Omega$cm. Interestingly, this is the range of $\rho$ where a metal-insulator transition occurs in many perovskite oxides~\cite{Arup95}.

The issue of magnetic transition at high fields (transition from the metallic to ferromagnetic metallic phase) is not very clear at this stage in the absence of high field magnetization data. We note that even at a field of 10T in the vicinity of 250K the ratio $B/T$ is not large enough to produce substantial spin alignment. (We are dealing with a S=3/2 system which is the core $t_{2g}$ spin). As a result the metallic phase that arises from the PI is likely to be paramagnetic. At low temperatures in a magnetic field the ground state is a ferromagnetic metallic phase. There appears to be some kind of transition/crossover as can be seen from the resistivity data at 12T in the vicinity of 215K. This being a transition in a magnetic field can be classified as a field induced ferromagnetic state. To sum up, we propose the scenario that in the vicinity of the phase diagram around the point ($H^*,T^*$) and at higher temperatures, the magnetic field increases the bandwidth and this suppresses the polaronic nature of the transport and any local charge or orbital order. This lowers $T_{CO}$ and at high enough field no charge or orbital order forms leading to a metallic phase. At still higher field there is field induced ferromagnetic state as the sample is cooled.

In conclusion, using calorimetric and transport investigations, we have shown that beyond a certain magnetic field, the COI state becomes unstable at all T.  The stability of the COI state is ensured by a finite $\Delta S_{CO}$. When $\Delta S_{CO}$ $\rightarrow$ 0, on the application of a magnetic field the COI state becomes unstable.

\section{Acknowledgment}AKR acknowledges financial support from DST (Government of India) for a sponsored project and KSN acknowledges CSIR for a fellowship.


\begin{references}
\bibitem{note1}Note:The COI-FMM transition in polycrystalline $Pr_{0.8}Na_{0.2}MnO_3$ have been investigated~\cite{prna}. application of magnetic field of 14T suppresses the specific heat associated with $T_{CO}$. However, the issue of entropy at transition has not been discussed. also the authors did not find any metallic nature at high field presumably due to polycrystalline nature of the material.
\end{references}
\end{document}